\documentclass{article}

     \PassOptionsToPackage{numbers, compress}{natbib}

\usepackage[final]{neurips_2024_ml4ps}

\usepackage{subfigure}
\usepackage{graphicx}

\newcommand{\rbx}{\texttt{RUBIX}}
\usepackage{amsmath}
\usepackage[utf8]{inputenc} 
\usepackage[T1]{fontenc}    
\usepackage{hyperref}       
\usepackage{url}            
\usepackage{booktabs}       
\usepackage{amsfonts}       
\usepackage{nicefrac}       
\usepackage{microtype}      
\usepackage{xcolor}         
\usepackage{wrapfig}

\title{Fast GPU-Powered and Auto-Differentiable Forward Modeling of IFU Data Cubes}

 \author{%
   Ufuk Çakır\thanks{Corresponding Author: Now at Intelligent Earth UKRI Centre for Doctoral Training in AI for the Environment, University of Oxford } \\
   Interdisciplinary Center\\ for Scientific Computing,\\
   University of Heidelberg,\\
   Im Neuenheimer Feld 205,\\ D-69120 Heidelberg\\
   \texttt{mail@cakir-ufuk.de} \\
     \And
     Anna Lena Schaible\\ 
   Interdisciplinary Center\\ for Scientific Computing,\\  University of Heidelberg,\\ Im Neuenheimer Feld 205,\\ D-69120 Heidelberg\\
   \texttt{annalena.schaible@iwr.uni-heidelberg.de}\\
   \And
      Tobias Buck\\ 
   Interdisciplinary Center\\ for Scientific Computing,\\  University of Heidelberg,\\ Im Neuenheimer Feld 205,\\ D-69120 Heidelberg\\
   \texttt{tobias.buck@uni-heidelberg.de}
 }

\begin{document}

\maketitle

\begin{abstract}
We present \rbx{}, a fully tested, well-documented, and modular Open Source tool developed in JAX, designed to forward model IFU cubes of galaxies from cosmological hydrodynamical simulations. The code automatically parallelizes computations across multiple GPUs, demonstrating performance
improvements over state-of-the-art codes by a factor of 600. This
optimization reduces compute times from hours to only seconds. \rbx{} leverages JAX’s auto-differentiation capabilities to
enable not only forward modeling but also gradient computations
through the entire pipeline paving the way for new methodological approaches such as e.g. gradient-based optimization of astrophysics model parameters. \rbx{} is open-source and available on GitHub\footnote{\url{https://github.com/ufuk-cakir/rubix}}.
\end{abstract}

\section{Motivation}
In the field of astrophysics, researchers are divided into two main groups: observers and theorists. Observers build and operate advanced instruments and telescopes, such as the James Webb Space Telescope (JWST) and the Very Large Telescope (VLT), to collect empirical data from distant galaxies and stars by counting photons. Integral Field Unit (IFU) spectroscopy is one key observational technique that produces datacubes with spatially resolved spectra.
%
Theorists, on the other hand, develop and refine physical equations to model the Universe's behavior. They use high-end supercomputers to run cosmological simulations, to replicate the conditions of the early universe. These simulations help test the implications of various physical theories and require advanced computational techniques and statistical analysis.
%
%
One significant challenge in astrophysics is bridging the gap between observational data and theoretical models. 
Forward modeling techniques, which translate simulation outputs into observable data, are crucial for effective collaboration.
The advances of Machine Learning models are hugely influenced by the improvements of hardware that has an architecture that works well with the calculations performed in ML applications: the GPU. They are extremely well suited to perform calculations in parallel, hence implementing a IFU forward model code that works on GPUs is a major advantage to current state-of-the-art codes and will enable us to produce a sufficient number of samples required for statistical analysis, which was so far the bottleneck for Machine Learning (ML) applications. Additionally, with our JAX implementation we are able to compute gradients needed to perform optimization in the context of ML and Simulation Based Inference.
%
%
This paper aims to bridge the gap between observers and theorists by introducing a forward modelling of mock IFUs: \rbx{} is written in JAX, runs natively parallel on multiple GPUs and leverages performance improvements from just-in-time compilation using XLA.

\section{Related Work}
\label{related work section}
There are several codes that can forward model IFU data that are commonly used in astrophysics research. One of the most popular codes is SimSpin \citep{SimSpin}, which is written in R and is a CPU only package that takes in a simulation of a galaxy and produces mock IFU observations. The user can freely choose any instrument configuration and spectral library to create mock observations. There is extensive documentation
and examples available, which makes the usage of the package very user-friendly.
Another code called GalCraft \citep{galcraft_paper} generates mock IFU data cubes of the Milky Way (MW). GalCraft uses the mock stellar catalog that is based on the analytical chemodynamical model of \cite*{sharma_2021_b}. The analytical model predicts the joint distribution of position, velocity, age, extinction, photometric magnitude and the chemical abundances of stars in the MW, which is then used to produce mock IFU data cubes.
In \cite*{Sarmiento_2023_MANGIA}, the authors emulated observation data from the MaNGA survey \citep{manga_paper} using IllustrisTNG data \citep{Pillepich_2019} to generate mock observations. Previously, \cite*{Bottrell_2022_Real_sim_ifs} emulated 893 MaNGA observations using the \textit{RealSimIFS} code from data of the TNG50 simulation \citep{Pillepich_2019, Nelson_2019}. A similar project of \cite*{nanni_2022_iManga} produced a catalog of around 1000 unique mock IFU observation to mimic the MaNGA primary sample -- again using data from the TNG50 simulation.
However, all current codes are CPU only use and have no option of calculating the gradient of the forward modelling process with respect to the input parameters - hence limiting their applicability within the context of ML.

\section{RUBIX Codebase}
The \rbx{} pipeline is a modular and efficient framework for forward modeling IFU data from cosmological simulations, leveraging the power of JAX for high-performance computing. \rbx{} is implemented as a linear pipeline, where each function sequentially transforms the input data, ensuring that the framework remains extremely modular and easily extensible for future developments. \rbx{} utilizes multiple GPUs for parallel computation and significantly reduces processing time.
The Input Handler extracts and transforms relevant star and gas particle information from cosmological simulation data into a unique data file, which is the input for the \rbx{} pipeline.
%
%
The first step in the pipeline is to orientate the galaxy in the field of view, following the specifications provided in the configuration file. Next, the particles are assigned to the IFU spaxels, accounting for telescope-specific configurations. The key part is the spectra calculation. Each star spectrum is calculated as a lookup from a simple stellar population (SSP) library. Then the spectra are Doppler shifted based on the galaxy distance and line-of-sight velocity of each stellar particle. Additional resampling is performed to match the wavelength grid of the observed telescope. Afterwards, the stellar spectra in each spaxel are summed up. To simulate observational effects, we apply a point-spread function (PSF) and line-spread function (LSF) convolution and add realistic noise.
%
%
%
%
%
%
%
%
%
The entire pipeline is configurable using JSON. Each pipeline run starts with a JSON or Python dictionary, where the user chooses all the hyperparameters, i.e SSP library, galaxy distance and orientation, telescope, etc. 

\section{Results}
\label{Results}
\paragraph{Qualitative Analysis}

\begin{figure}
    \centering
    \begin{minipage}{0.49\linewidth}
        \centering
        \includegraphics[width=\linewidth,trim=0 0 0 0, clip]{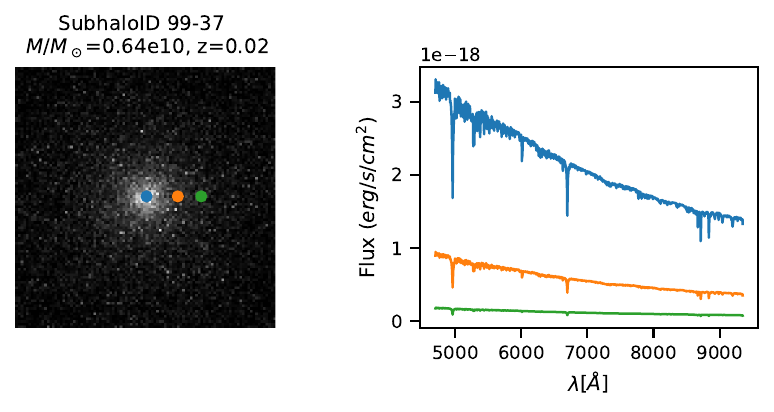}
        \label{fig:subfig1}
    \end{minipage}
    \hfill
    \begin{minipage}{0.49\linewidth}
        \centering
        \includegraphics[width=\linewidth,trim=0 0 0 0, clip]{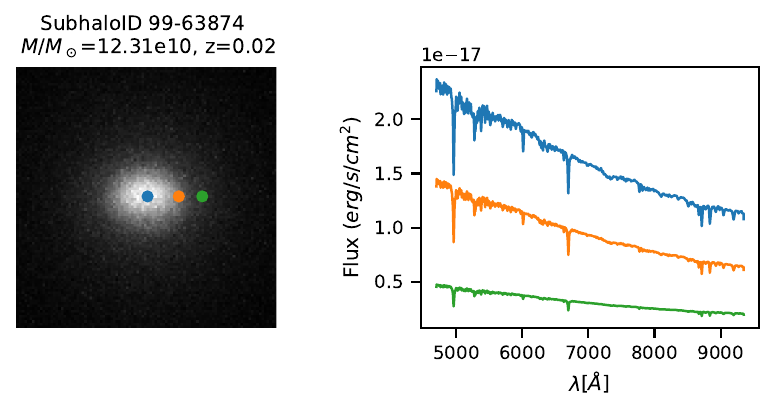}
        \label{fig:subfig4}
    \end{minipage}
    \vspace{-.5cm}
    \caption{\textbf{\textit{MUSE mock observations}} -- for different Subhalos, the total flux in each pixel is shown as an image representation on the left. On the right, the spectra of three different spaxels are plotted.}
    \label{fig:muse-mock-observations}
\end{figure}
To verify the output, \rbx{} is executed for different Subhalos from the IllustrisTNG simulation using the IllustrisAPI. For a set of galaxies from the TNG50-1 simulation, snapshot 99, mock observations are created with a MUSE instrument configured with \texttt{fov=20} and a Gaussian PSF and LSF. The \texttt{Mastar\_CB19\_SLOG\_1\_5} SSP template \cite{Sánchez_2022} is employed to compute the stellar spectra, and \rbx{} is executed on eight NVIDIA A100 GPUs.
The mock MUSE observations are illustrated in Figure \ref{fig:muse-mock-observations}. 
The galaxies were chosen to have an increasing mass, with a dwarf galaxy on the left and a massive spherical galaxies on the right. For each galaxy, an image representation is provided on the left, where the total flux in each spaxel is summed to produce a two-dimensional array. The right column presents the spectra (in units of erg/s/cm²) for three different spaxels with increasing distance from the galactic center.
Spectra from the center of the galaxy exhibit higher flux compared to those from the outskirts, which is expected due to the higher density of stars in the galaxy's center. The shapes of the spectra differ significantly; spectra from the outskirts tend to be flatter. In general we can observe that \rbx{} can reproduce the trends that we expect.




\begin{wrapfigure}[16]{R}{0.5\textwidth}
   \vspace{-0.6cm}
   \centering
   \includegraphics[width=0.49\textwidth,trim={0cm 0cm 0cm 0.8cm},clip]{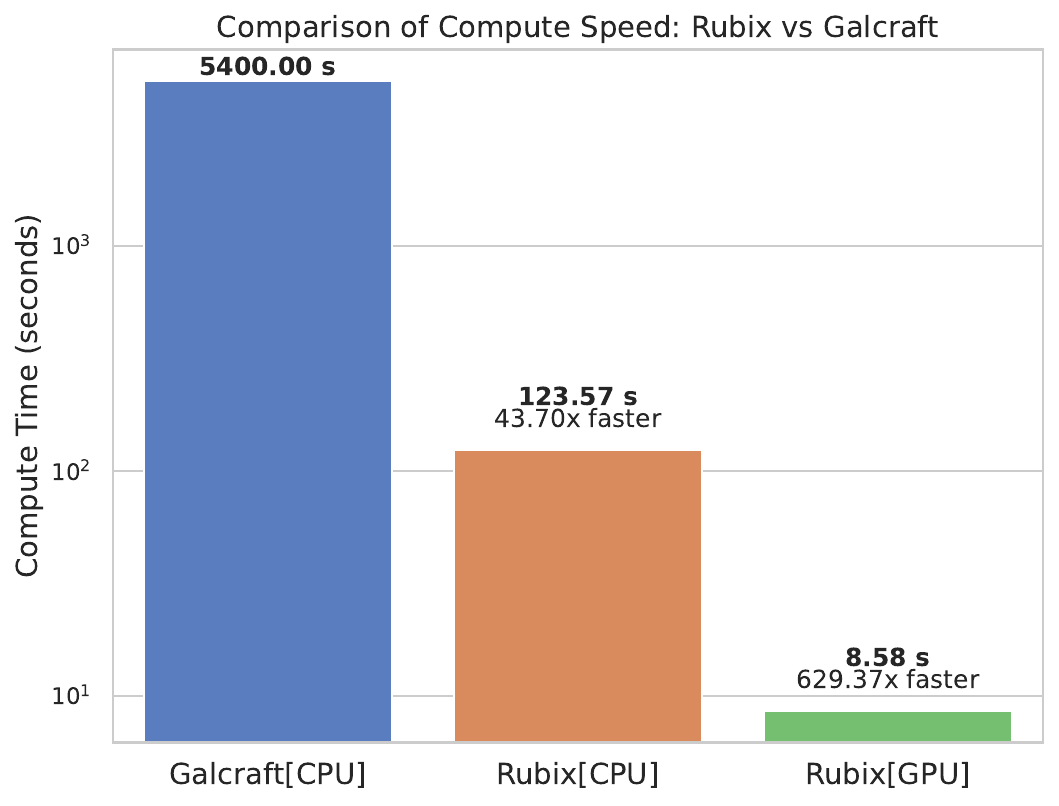}
   \vspace{-.25cm}
   \caption{\textbf{\textit{Speed comparison}} -- the execution time of different codes are compared. Note that the y-axis is logarithmic.}
   \label{fig:rubix_galcraft_speed_comparison}
\end{wrapfigure}
\paragraph{Speed comparison}
The primary objective of this paper is to highlight the methodological improvements \rbx{} provides for the forward modeling process. In Figure \ref{fig:rubix_galcraft_speed_comparison}, the compute times of different codes are compared. According to \cite*{galcraft_paper}, the authors state that "for a typical MUSE FoV containing $6\cdot 10^6$ particles, the execution time spent with a 24-core CPU (2.50GHz) is 1.4 hours." This result serves as a benchmark to contextualize the execution time of \rbx{}.
A galaxy with a comparable number of particles ($6\cdot 10^6$) is forward modeled both on the CPU and GPU using \rbx{}. Running \rbx{} on a 24-core CPU (AMD Epyc 7452, 2.35 GHz) takes 123.57 seconds, representing a 43.7-fold improvement over the Galcraft code. When the mock observation is computed on a single NVIDIA A-100 GPU, the execution time is reduced to 8.58 seconds, which is 600 times faster than Galcraft and 14.4 times faster than the same \rbx{} code executed on the CPU.
Despite the clear performance improvements, we should take the benchmark comparisons with caution, because we use different hardware configurations. GPUs and CPUs have different architecture and GalCraft does not share the exact same methodology. Despite these differences, the comparison still offers a valuable general trend. One significant reason for \rbx{}'s superior speed is its efficient implementation. In \rbx{}, instead of naively looping over the particles, every function is vectorized using \texttt{vmap}. This approach leverages XLA to fuse operations together, resulting in substantial speed improvements.

\begin{figure}
    \centering
        \subfigure[\textbf{\textit{Strong Scaling}}]{\includegraphics[width=.32\textwidth]{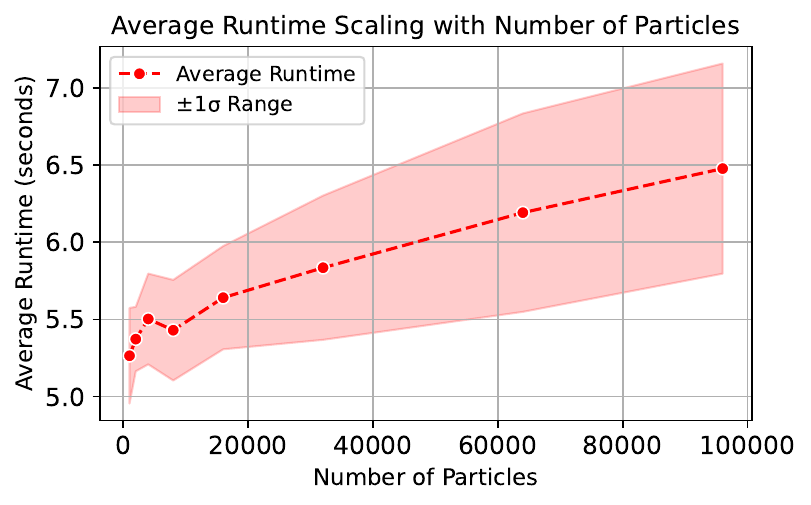}
        \label{fig:strong-scaling}} 
        \subfigure[\textbf{\textit{Weak Scaling}}]{\includegraphics[width=.32\textwidth]{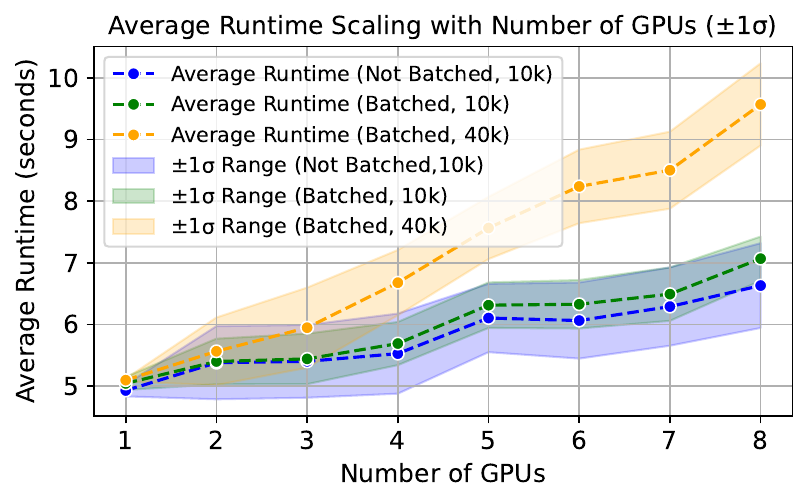}
        \label{fig:weak-scaling}}
        \subfigure[\textbf{\textit{Scaling Efficiency}}]{\includegraphics[width=.32\textwidth]{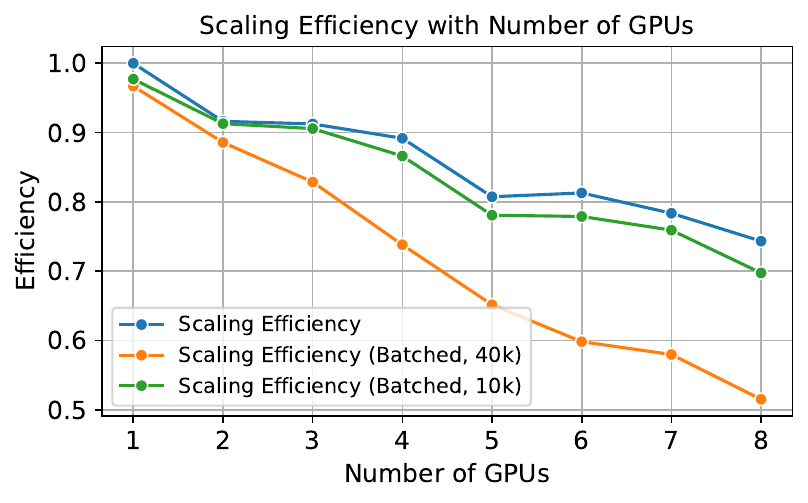}
        \label{fig:scaling-efficiency}}
    \caption{\textbf{\textit{Scaling plots}} --
    (a) Strong Scaling: Increasing particle size, while keeping number of GPUs fixed (8 NVIDIA A100 GPUs).
    (b) Average runtime of different \rbx{} runs, where we proportionally increase particle size and number of available GPUs, such that the workload per GPU remains constant.
    (c) Scaling efficiency calculated as the ratio of the runtime with one GPU to the runtime with multiple GPUs.}
    \label{fig:combined-weak-scaling-efficiency}
\end{figure}

\paragraph{Strong Scaling}
In Figure \ref{fig:strong-scaling}, the average runtime is plotted against the number of particles. At each number of particles, the runtime is measured five times to get some statistics. The red shaded area represents the ±1$\sigma$ range, indicating the variability in the runtime measurements. One can observe that as the number of particles increases, the average runtime also increases, but not linearly.  This indicates that \rbx{} does not have perfect strong scaling, which may be caused by the communication overhead between the GPUs.

\paragraph{Weak Scaling}
To evaluate how compute time scales with the number of GPUs, \rbx{} is initially run with 10,000 particles on a single GPU, and the runtime is measured. Next, the number of particles is doubled, and the code is executed on two GPUs, continuing this process until the maximum number of GPUs is reached, which in this case is eight NVIDIA A100 GPUs. In Figure \ref{fig:weak-scaling}, the average runtime of different \rbx{} runs is measured. Each run is repeated 5 times to get some statistics, and the 1$\sigma$ area is shaded in the background (blue: starts with 10,000 particles; green: split data into four batches on each GPU; orange: starts with 40,000 particles and batching). 
Ideally, in a best-case scenario, the compute time should remain constant as both the workload (number of particles) and the compute resources (number of GPUs) increase proportionally. This would demonstrate perfect scaling. From Figure \ref{fig:weak-scaling} we can clearly see that the scaling is not perfect. The runtime increases slowly with the number of GPUs, indicating that the computational work is not distributed equally between the GPUs or that communication overheads are still large in \rbx{}.

\paragraph{Scaling Efficiency}
To make this more quantitatively, we can measure the scaling efficiency as:
 $   \text{Scaling Efficiency} = \frac{T_1}{T_N}$
where $T_1$ is the runtime with one GPU and $T_N$ is the runtime with $N$ GPUs. Ideally, this scaling efficiency should be close to one. In Figure \ref{fig:scaling-efficiency} we clearly see that the scaling efficiency decreases with increasing number of GPUs, which means that the scaling is not optimal. One major factor is the communication overhead between GPUs, which can become significant as more GPUs are added. Additionally, the efficiency of load balancing can decrease with more GPUs. Furthermore, the complexity of managing more GPUs can introduce inefficiencies in the parallelization process, such as increased latency in coordinating tasks and distributing data evenly among the GPUs.
%
%
%
%
This indicates that \rbx{} is not yet fully optimized and requires further improvements. One significant bottleneck might be the current implementation of \texttt{pmap} and \texttt{jit}. In the current version of \rbx{}, only the datacube calculation inside the pipeline is parallelized across the GPUs using \texttt{pmap}. However, during the pipeline assembly, all the functions are concatenated, and the final function is just-in-time compiled using \texttt{jit}. There is a known issue in JAX warning users that using \texttt{jit} on a \texttt{pmap}-function can lead to inefficient data movement, as it essentially collects all data onto a single device. This issue is discussed in detail on the official JAX GitHub page\footnote{\url{https://github.com/google/jax/issues/2926}}.

\section{Conclusions and Limitations}
\rbx{} represents a significant leap forward in computational efficiency and flexibility for modeling IFU
observations from cosmological hydrodynamical simulations. Its ability to rapidly process large-scale simulations and its potential for future enhancements makes it a powerful tool for astrophysical research. The combination of high performance and open-source accessibility underscores the contribution of \rbx{} to the field, facilitating innovation and collaboration within the scientific
community.
Despite its impressive performance, there remains potential for further optimization, i.e. further profiling is required. 
Apart from speed improvements, there are additional features that will be implemented into \rbx{}. Some of those include:

\begin{itemize}
    \item \textbf{Gas Modeling} -- Incorporating detailed models for interstellar gas will allow \rbx{} to simulate and analyze gas emission lines. 
    \item \textbf{Dust Modeling} -- Adding support for dust attenuation models will provide more realistic mock observations, that should closer relate to real observations.
    \item \textbf{Radiative Transfer} -- Implementing advanced radiative transfer models will enhance the precision of the \rbx{} simulations. This will allow for a more realistic representation of how light propagates through various media. However, this needs to be implemented in pure JAX, which can be a quite challenging task.
\end{itemize}

With these additional features \rbx{} will be ideally suited to tackle key scientific machine learning tasks in astrophysics, such as performing SBI inference of fundamental galaxy parameters with high-dimensional complex observational data, perform Bayesian model comparison, do gradient based optimization tasks on the forward modelling pipeline and incorporate the differentiable forward model \rbx{} into machine learning architectures to train them end-to-end, e.g. build hybrid NN encoder-physics-based-decoder architectures. As such, we think that \rbx{} provides the astrophysical community with a unique, versatile and new methodological approach to perform downstream scientific tasks.


\section*{Broader impact statement}
The authors are not aware of any immediate ethical or societal implications of this work. This work purely aims to aid scientific research and proposes a method of using a pipeline of forward modelling IFU data cubes to learn about galaxy formation and evolution.

\begin{ack}
The authors thank the Scientific Software Center at Heidelberg University for the support. This work is funded by the Carl-Zeiss-Stiftung through the NEXUS programm.
\end{ack}

\bibliographystyle{plain}
\bibliography{lit.bib}


\appendix

\end{document}